\documentclass[%
reprint,
superscriptaddress,
amsmath,amssymb,
aps
]{revtex4-2}

\usepackage{graphicx}
\usepackage[dvipsnames]{xcolor}
\usepackage{dcolumn}% Align table columns on decimal point
\usepackage{bm}% bold math
\usepackage[utf8]{inputenc}
\usepackage{siunitx}
\usepackage[T1]{fontenc}
\usepackage{amsmath}
\usepackage{multirow}

\begin{document}
%\preprint{APS/123-QED}
%\preprint{}

\title[Dong et al.]{Growth optimization of shallow Ge quantum wells grown by molecular beam epitaxy}

\author{J. T. Dong}
\email[]{jtdong@lps.umd.edu}
\affiliation{Laboratory for Physical Sciences, 8050 Greenmead Drive, College Park, MD 20740}
\affiliation{Department of Physics, University of Maryland, College Park, MD 20742}

\author{J. P. Thompson}
\affiliation{Laboratory for Physical Sciences, 8050 Greenmead Drive, College Park, MD 20740}
\affiliation{Department of Physics, University of Maryland, College Park, MD 20742}

\author{R. Card}
%\email[]{}
%\thanks{}
\affiliation{Laboratory for Physical Sciences, 8050 Greenmead Drive, College Park, MD 20740}
\affiliation{Department of Materials Science \& Engineering, University of Maryland, College Park, MD 20742}

\author{K. Sardashti}
%\email[]{ksardash@umd.edu}
\affiliation{Department of Physics, University of Maryland, College Park, MD 20742}
\affiliation{Laboratory for Physical Sciences, 8050 Greenmead Drive, College Park, MD 20740}

\author{C. J. K. Richardson}
%\email[]{richardson@lps.umd.edu}
\affiliation{Laboratory for Physical Sciences, 8050 Greenmead Drive, College Park, MD 20740}
\affiliation{Department of Materials Science \& Engineering, University of Maryland, College Park, MD 20742}

\date{\today}

\begin{abstract}
Shallow strained Ge quantum wells have gained recent attention for realizing scalable, high-performance hybrid superconductor/semiconductor-based qubits. Epitaxial superconducting contacts can improve the quality and consistency of the superconductor/semiconductor interface. The growth of Ge quantum wells by molecular beam epitaxy is then motivating due to the relative ease of integration with epitaxial superconductor growth. However, the performance of Ge quantum wells grown by molecular beam epitaxy (MBE) has been limited. To improve the properties of MBE-grown Ge quantum wells, the growth conditions were systematically optimized. Thick buffer layers are utilized to eliminate certain defects, and an optimal growth temperature is found. A peak hole mobility of 105,000 cm\textsuperscript{2}/Vs at 2 K is obtained in a 22-nm deep Ge quantum well, demonstrating that the Ge quantum wells grown in this study represent the highest mobilities for shallow MBE-grown samples. Mobility modeling indicates that the increase in mobility due to growth temperature optimization are likely due to a reduction in interface roughness scattering, and further improvements in mobility are expected through improved surface passivation.

\end{abstract}


\maketitle %\maketitle must follow title, authors, abstract and \pacs

\section*{Introduction}

% Do a super/semi intro?

Strained Ge quantum wells (QWs) are an emerging material system for solid-sate quantum computing. Ge QWs are the leading platform for hole-based spin qubits \cite{hendrickx_single-hole_2020,hendrickx_fast_2020,hendrickx_sweet-spot_2024,van_riggelen-doelman_coherent_2024}, and shallow Ge QWs have also been recently investigated for hybrid superconductor/semiconductor qubits \cite{sagi_gate_2024,kiyooka_gatemon_2025}. In this application, Ge QWs are a particularly promising platform to realize improved hybrid qubits due to the potential for monolithic integration of low-loss superconducting circuitry on Si substrates with Ge QW-based Josephson junctions or spin qubits.

Excellent transport properties have been demonstrated in chemical vapor deposition (CVD)-grown Ge QWs \cite{Sammak2019,zhang_sharp_2022,kong_undoped_2023,Myronov2023,Stehouwer2023}, and CVD is typically used to grow high-quality Ge QWs. Due to the difficulty of integrating CVD with ultrahigh vacuum-based deposition techniques commonly used for high-quality superconductor growth, all demonstrations of Ge QW-based Josephson junctions have been limited to \textit{ex-situ} deposited superconducting contacts \cite{vigneau_germanium_2019,tosato_hard_2023,valentini_parity-conserving_2024}. Issues such as low and inconsistent transparencies (quantified through the I\textsubscript{C}R\textsubscript{N} product) are reported in these devices, and are likely due to residual contamination at the superconductor/semiconductor interface from the \textit{ex-situ} processing. 

The growth of Ge QWs by molecular beam epitaxy (MBE) is therefore motivating to enable integration of epitaxial superconducting contacts with consistent properties and high transparency. However, there are few studies on the growth of Ge QWs by MBE, and the transport properties of the MBE-grown Ge QWs are often lacking in comparison to CVD-grown QWs \cite{Xie1993,sigle_strained_2021}. The highest reported mobility in a MBE-grown Ge QWs is 120,000 cm\textsuperscript{2}/Vs in a 35-nm deep QW \cite{zhang_high-quality_2024}, which is a value over an order of magnitude lower than the values reported in similar CVD-grown Ge QWs \cite{kong_undoped_2023}. To further improve the mobility of MBE-grown Ge QWs, a thorough investigation of the growth conditions is necessary. In this work, the optimization of the Ge QW growth by MBE is investigated. Significant improvements to the low temperature mobilities are observed, and the effect of varying the growth conditions on the scattering of holes in the Ge QW is studied. Further improvements to the mobility of MBE-grown Ge QWs will be discussed.

\section*{Methods}
\begin{figure}[t]
\centering
\includegraphics[]{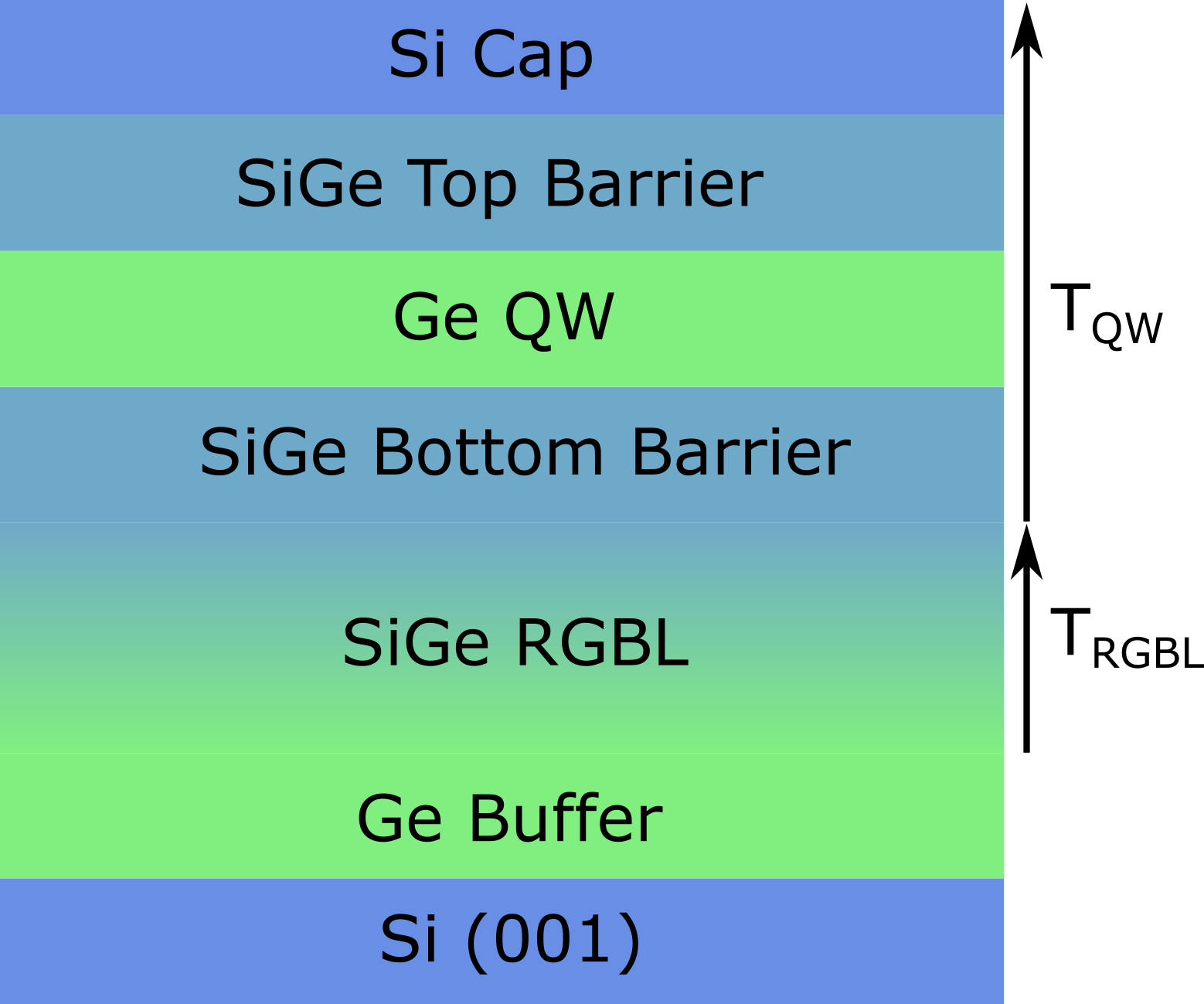}
\caption{\label{fig:1} Schematic layer structure of the Ge QWs. Layer thicknesses are not shown to scale}
\end{figure}

Ge QWs were grown by MBE on high resistivity float-zone Si (001) wafers. Ge and Si electron beam evaporators were used as the Ge and Si sources. Before loading into the deposition system, the Si wafers were cleaned with a 5\% HF etch to remove the native oxide on the surface. After loading into the system, the Si wafers were annealed at 800~\unit{\degreeCelsius} to desorb residual oxides on the surface. A 500-nm-thick Ge buffer was subsequently grown, using a two step growth technique to reduce the threading dislocation density \cite{shah_high_2011}. The first 100 nm were grown at 300~\unit{\degreeCelsius}, and the remaining thickness were grown at 550~\unit{\degreeCelsius}. The Ge buffer was subsequently annealed at 850~\unit{\degreeCelsius} to further reduce the threading dislocation density. Once the anneal is completed, a reverse graded buffer layer (RGBL) was grown with the composition of the RGBL linearly graded from Ge to Si\textsubscript{0.2}Ge\textsubscript{0.8}. A 400-nm-thick, relaxed Si\textsubscript{0.2}Ge\textsubscript{0.8} bottom barrier, a 16-nm-thick compressively strained Ge QW, a 22-nm-thick relaxed Si\textsubscript{0.2}Ge\textsubscript{0.8} top spacer, and a 1-nm-thick Si cap were grown. A schematic layer heterostructure is shown in Fig. \ref{fig:1}. The growth temperature of the bottom barrier, the quantum well, the top spacer, and the cap are defined as the QW growth temperature for this study. In this study, different RGBL thicknesses (t\textsubscript{RGBL}), RGBL growth temperatures (T\textsubscript{RGBL}), and QW growth temperatures (T\textsubscript{QW}) were investigated. The parameters explored in this study are provided in Table \ref{tab:1}.

\begin{table*}[]
\caption{\label{tab:1}
Ge QW heterostructure layer composition, thicknesses, and growth temperatures explored in this study}
\begin{ruledtabular}
\begin{tabular}{ l c c c c c c c}
%    & Ge & SiGe & SiGe & Ge & SiGe & Si \\ 
  & Substrate &  Buffer &  Reverse Graded Buffer & Bottom Barrier & QW & Top Barrier & Cap \\ 
\hline
Composition & Si & Ge & Ge $\rightarrow$ Si\textsubscript{0.2}Ge\textsubscript{0.8} & Si\textsubscript{0.2}Ge\textsubscript{0.8} & Ge & Si\textsubscript{0.2}Ge\textsubscript{0.8} & Si \\
Thickness (nm) & -- & 100/400 & 700 -- 2000 & 400 & 16 & 22 & 1 \\ 
Temperature (\unit{\degreeCelsius}) & -- & 300/550 & 300 -- 600 & \multicolumn{4}{c}{300 -- 575} \\ 
\end{tabular}
\end{ruledtabular}
\end{table*}

The samples were etched with defect-selective iodine based solution consisting of HF:HNO\textsubscript{3}:CH\textsubscript{3}COOH:I\textsubscript{2} in a ratio of 5:9:10:1, which is further diluted 1:1 with water and the samples were etched for 11 s to reveal structural defects such as threading dislocations and stacking faults. This etch results in an approximate etch depth of 100 nm, revealing structural defects within the SiGe bottom barrier. The top surface roughness and morphology was characterized with atomic force microscopy (AFM).

\begin{figure*}[]
\centering
\includegraphics[]{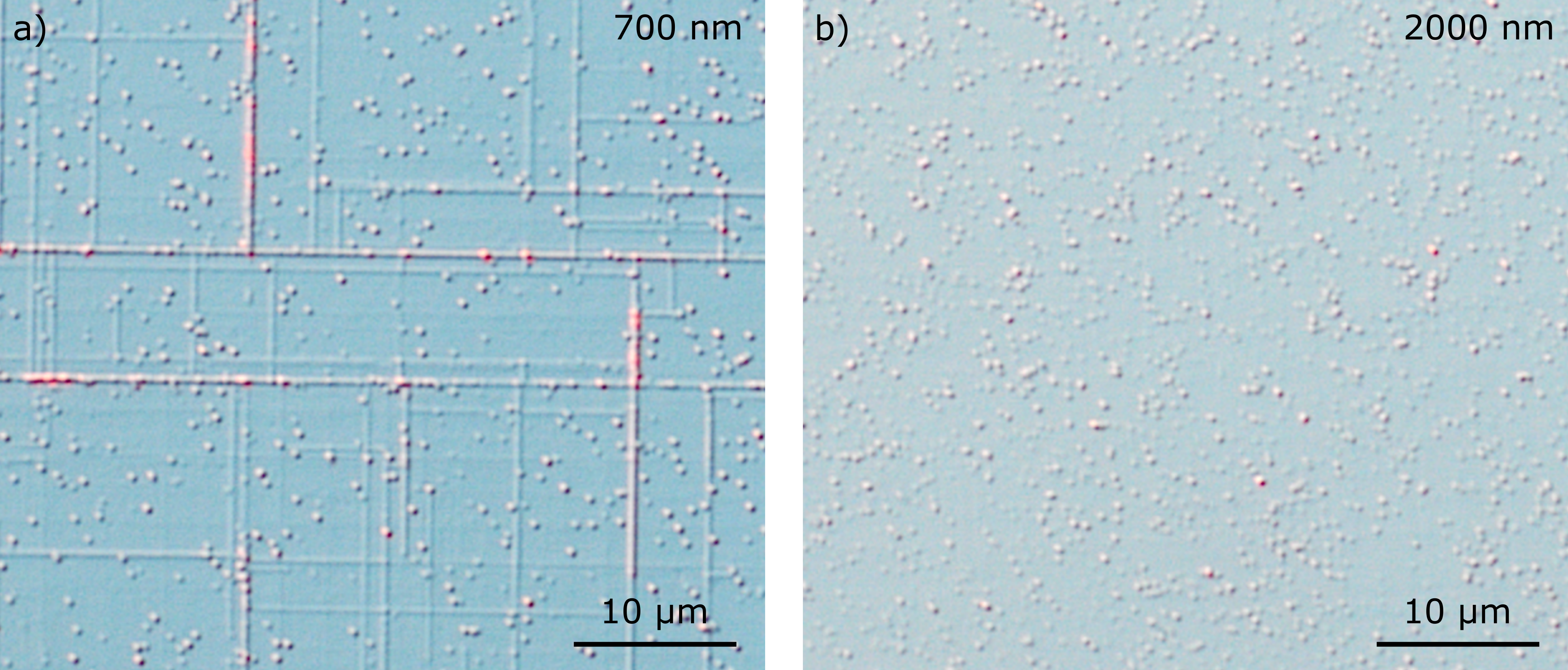}
\caption{\label{fig:2} Post iodine etch Nomarski micrographs of Ge QWs grown with  a) 700-nm-thick reverse graded buffer and b) 2000-nm-thick reverse graded buffer. Both are grown at 300~\unit{\degreeCelsius}.}
\end{figure*}
% Shorten figure?

Gated Hall bars were fabricated to measure the mobility of the different samples. To form an Ohmic contact to the Ge QWs, 50-nm-thick Pt contacts were deposited with an electron-beam evaporator. The QWs were etched with a buffered oxide etch for 10 s prior to Pt deposition. The Pt contacts were annealed at 300 \unit{\degreeCelsius} for 1 hour in an Ar environment to form the Ohmic contacts to the Ge QWs. The Hall bar mesa was etched with a BCl\textsubscript{3}-based reactive ion etch. Before gate dielectric deposition, the samples were exposed to an O\textsubscript{2} plasma to passivate the surface. A 30-nm-thick AlO\textsubscript{x} gate dielectric was deposited at 200 \unit{\degreeCelsius} with atomic layer deposition. Ti/Au was used as the gate electrode. The ratio of the Hall bar width to the spacing of the arms was 10. The Hall bars were cooled to 2 K in a Quantum Design Physical Property Measurement System and standard low-frequency lock-in techniques were used to measure the magnetotransport. The sheet resistance was determined from the zero magnetic field longitudinal resistance, while the carrier density was determined from a linear fit of the low magnetic field Hall resistance. All samples had single carrier transport devoid of parallel conduction. The mobility was then calculated from the sheet resistance and carrier density. The carrier density in the Hall bar was varied using different applied voltages to the gate electrode.

\section*{Results and discussion}

Structural defects within the material were identified with the iodine-based etch. A Nomarski micrograph of the defect-selective etched Ge QWs grown with T\textsubscript{RGBL}~=~300 \unit{\degreeCelsius} and varying t\textsubscript{RGBL} is shown in Fig. \ref{fig:2}. Etch pits due to threading dislocations are visible. The etch pit density of the Ge QWs is typically $\sim 5 \times 10^{7}$ cm\textsuperscript{-2}, which is comparable to values reported in CVD-grown Ge QWs \cite{Sammak2019,zhang_sharp_2022}. Line-shaped defects are present in the sample with a RGBL thickness of 700 nm. Similar line-shaped defects have previously been attributed to stacking faults, which can form due to contamination at the substrate/film interface \cite{shiraki_elimination_1975} or growth conditions of the SiGe buffer layers \cite{shah_reverse_2010}. Defect selective etching of the various layers of the heterostructure reveal that the line-shaped defects originate from the SiGe layers, indicating that the defects nucleate due to the RGBL growth conditions. Increasing the T\textsubscript{RGBL} from~300~--~600~\unit{\degreeCelsius} for t\textsubscript{RGBL} = 700 nm was ineffective at eliminating the line-shaped defects. Temperatures exceeding 600 \unit{\degreeCelsius} were not explored due to pitting formation for samples grown at elevated temperatures. Instead, increasing the t\textsubscript{RGBL} from 700 nm to thickness exceeding 1500 nm was found to eliminate the line-shaped defects. A small temperature dependence on the thickness that eliminates the line-shaped defect is observed, where thicknesses $\geq$ 2000 nm are required to eliminate the line-shaped defects for T\textsubscript{RGBL}~=~300 \unit{\degreeCelsius}, while thicknesses $\geq$ 1500 nm are sufficient to eliminate the line-shaped defects for T\textsubscript{RGBL}~=~600 \unit{\degreeCelsius}. These observations of a critical RGBL thickness for the elimination of the line defects is consistent with reports on CVD-grown RGBL requiring a sufficient thickness to inhibit stacking fault formation \cite{shah_reverse_2010}. However, in the CVD-grown RGBL a critical RGBL thickness of 200 nm was observed for RGBLs grown at 850 \unit{\degreeCelsius}. The large disparity in the critical thickness for eliminating line-shaped defects in RGBL between the CVD-grown and MBE-grown RGBLs is likely due to the significant growth temperature differences between these two growth techniques, and is consistent with the observed trend of higher growth temperatures requiring thinner RGBL thickness.

\begin{figure*}[t]
\centering
\includegraphics[]{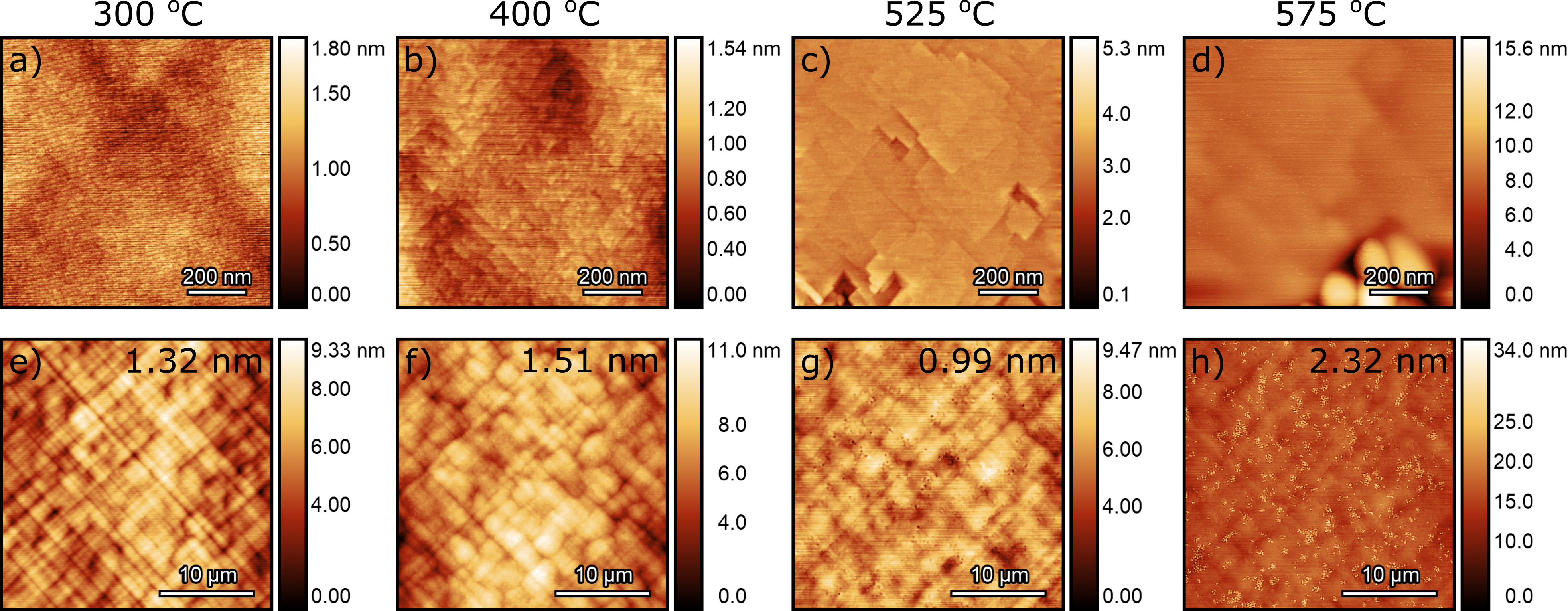}
\caption{\label{fig:3} a)-d) $1 \times 1$ \unit{\micro\meter} and e)-h) $30 \times 30$ \unit{\micro\meter} AFM micrographs and of the Ge QW grown at different temperatures as shown at the top of reach column. The root-mean-square roughness is given in e)-h).}
\end{figure*}

With the line-shaped defects in the heterostructures eliminated, the effect of varying the QW growth temperature was explored in Ge QWs grown with t\textsubscript{RGBL} = 2000 nm and the T\textsubscript{QW} = T\textsubscript{RGBL}. AFM micrographs of the top surface morphology with the varying growth temperature are shown in Fig. \ref{fig:3}. From the $1 \times 1$ \unit{\micro\meter} scans, a relatively featureless surface is observed for growths at 300 \unit{\degreeCelsius} and atomic steps become visible on the surface for T\textsubscript{QW}~$\geq$~400~\unit{\degreeCelsius}. As the temperature continues to increase, the spacing between steps increases. These changes in surface morphology are due to increased adatom diffusivity at higher temperatures, with a change in growth mode from layer-by-layer to step-flow at T\textsubscript{QW}~$\approx~400$~\unit{\degreeCelsius}. In the $30 \times 30$ \unit{\micro\meter} scans, a crosshatched surface morphology typical of metamorphic SiGe growth is present. As the T\textsubscript{QW} increases the cross hatch density decreases. Deep pits are present in the Ge QW grown at 575 \unit{\degreeCelsius}, which are likely due to thermal roughening of SiGe at high temperatures \cite{bean_getextsubscriptxsitextsubscript1-xsi_1984}. The RMS roughness of the Ge QWs from the $30 \times 30$ \unit{\micro\meter} remains $\sim$ 1.3 nm across the various growth temperatures, until T\textsubscript{QW}~=~575 \unit{\degreeCelsius}. An increase in the roughness to values greater than 2 nm is observed due to the surface pitting for T\textsubscript{QW} = 575 \unit{\degreeCelsius}. The RMS roughness of $\sim$ 1.3 nm is comparable to roughness values of Ge QWs grown by CVD \cite{Sammak2019,kong_undoped_2023,zhang_sharp_2022,zhang_high-quality_2024}. These results indicate that the changes in growth temperature primarily modify the local length scale of roughness instead of significantly modifying the magnitude of roughness for T\textsubscript{QW} $<$ 575 \unit{\degreeCelsius}, with higher growth temperatures promoting a longer length scale of roughness.

\begin{figure*}[]
\centering
\includegraphics[]{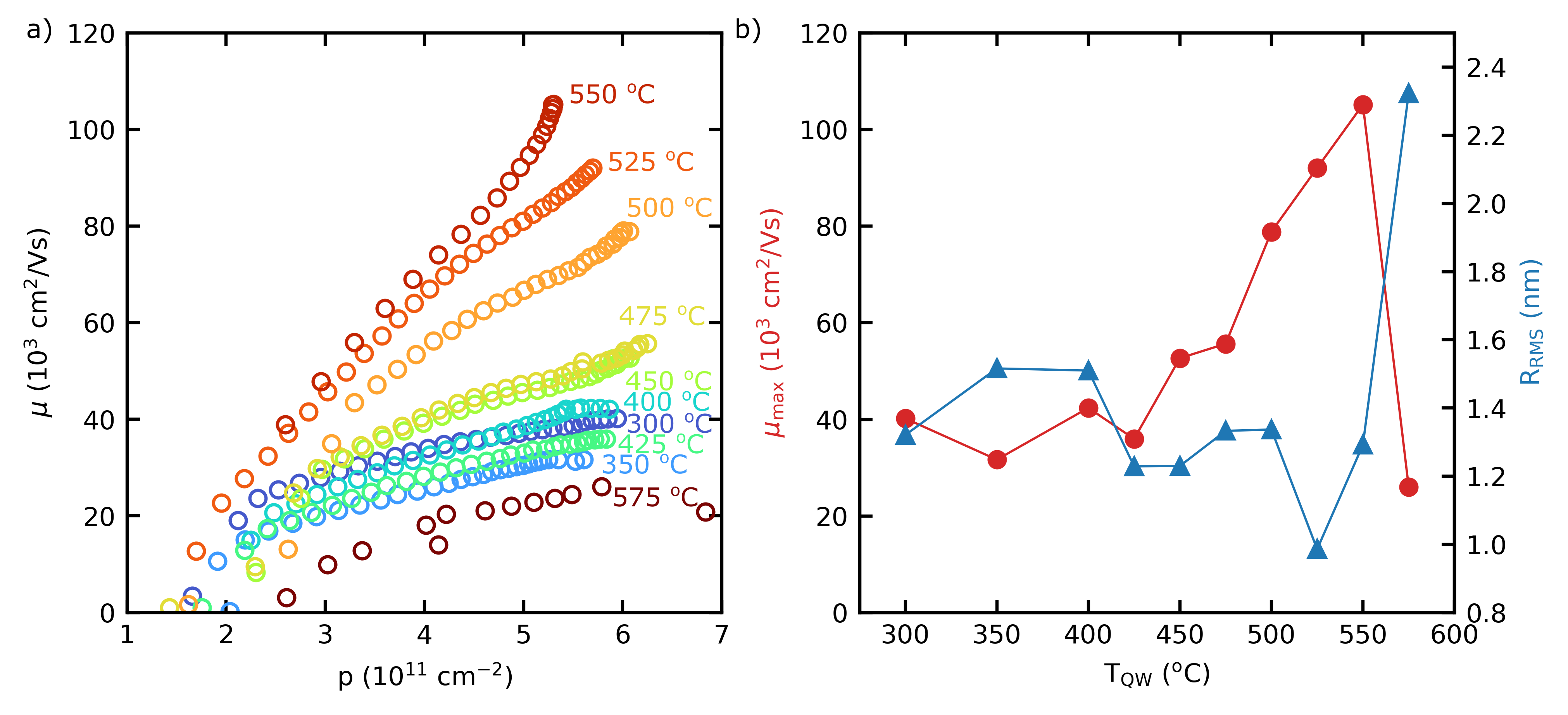}
\caption{\label{fig:5} a) Hole mobility dependence on carrier density for Ge QWs grown with varying temperatures, T\textsubscript{QW}. b) Maximum mobility and root-mean-square roughness at varying T\textsubscript{QW}. Gated Hall bar measurements performed at 2 K.}
\end{figure*}

The dependence of the 2 K mobility measured with the gated Hall bars fabricated from the Ge QWs grown with varying T\textsubscript{QW} is shown in Fig. \ref{fig:5} a). All Hall bars were in enhancement mode and a gate voltage is required to induce carriers in the Ge QW. All samples have a maximum induced carrier density that can be induced within the quantum well corresponding to a density of $\sim 5.5 \times10^{11}$ cm\textsuperscript{-2}, which is due to accumulation of carriers at the SiGe/oxide interface at excessive negative gate voltages and agrees well with the saturation carrier density of Ge QWs with similar depth from the surface \cite{Sammak2019,su_effects_2017}. The mobility in all samples typically increases with increasing carrier density as a result of greater screening of disorder and impurities from the higher carrier density. Higher mobility samples exhibit an increased slope of the dependence of mobility on carrier density. Increasing T\textsubscript{QW} typically increases the maximum mobility until T\textsubscript{QW} = 575 \unit{\degreeCelsius}, where a sharp drop in the maximum mobility is observed that is likely due to the pit formation and high roughness. The growths at T\textsubscript{QW} $<$ 450 \unit{\degreeCelsius} and at 575 \unit{\degreeCelsius} exhibit low mobilities and have similar slopes. Different scattering mechanisms have different characteristic slopes in the dependence of mobility on carrier density \cite{das_sarma_universal_2013}. The changes in slope observed in the differing samples indicate a change in the limiting scattering mechanisms of the samples due to the  growth conditions, with the low temperature growths and the growths at 575 \unit{\degreeCelsius} likely having similar limiting scattering mechanisms and the mobility being roughness limited. 

The maximum mobility and root-mean-square roughness of each sample is shown in Fig \ref{fig:5} b). From T\textsubscript{QW} = 300 - 425 \unit{\degreeCelsius} the mobility of the Ge QWs does not vary significantly from $\sim$ 40,000 cm\textsuperscript{2}/Vs. At growth temperatures above 425 \unit{\degreeCelsius}, the mobility begins to increase with increasing T\textsubscript{QW}. This onset of the mobility beginning to increase is correlated with the growth mode change from layer-by-layer to step flow, and suggests that the change in growth mode is related to the mobility improvements. At T\textsubscript{QW} above 450 \unit{\degreeCelsius}, increasing the growth temperature continues to increase the mobility until a maximum mobility of 105,000 cm\textsuperscript{2}/Vs is obtained for a Ge QW with T\textsubscript{QW} = 550 \unit{\degreeCelsius}. This mobility value is comparable to mobilities reported in significantly deeper Ge QWs grown by MBE \cite{zhang_high-quality_2024}, however it is a factor of 2 - 4 lower than the mobilities obtained in similar CVD-grown Ge QWs \cite{Sammak2019,Shimura2024}. These results demonstrate that these Ge QWs represent the highest mobilities for shallow MBE-grown Ge QWs, but more material improvements are required in order for MBE-grown Ge QWs to have mobilities similar to CVD-grown ones. The observed improvements in mobility are not correlated with changes in the root-mean-square roughness, suggesting that changes in the roughness are not responsible for the mobility enhancement. A further increase in growth temperature to 575 \unit{\degreeCelsius} causes a rapid degradation in mobility and coincides with the surface roughening. The highest growth temperature before roughening occurs is the optimal growth temperature for maximizing mobility. A narrow growth window for obtaining high mobility Ge QWs is present, and precise control over the growth temperature of the Ge QWs is required in order to reproducibly obtain the highest mobilities.

\begin{figure*}[]
\centering
\includegraphics[]{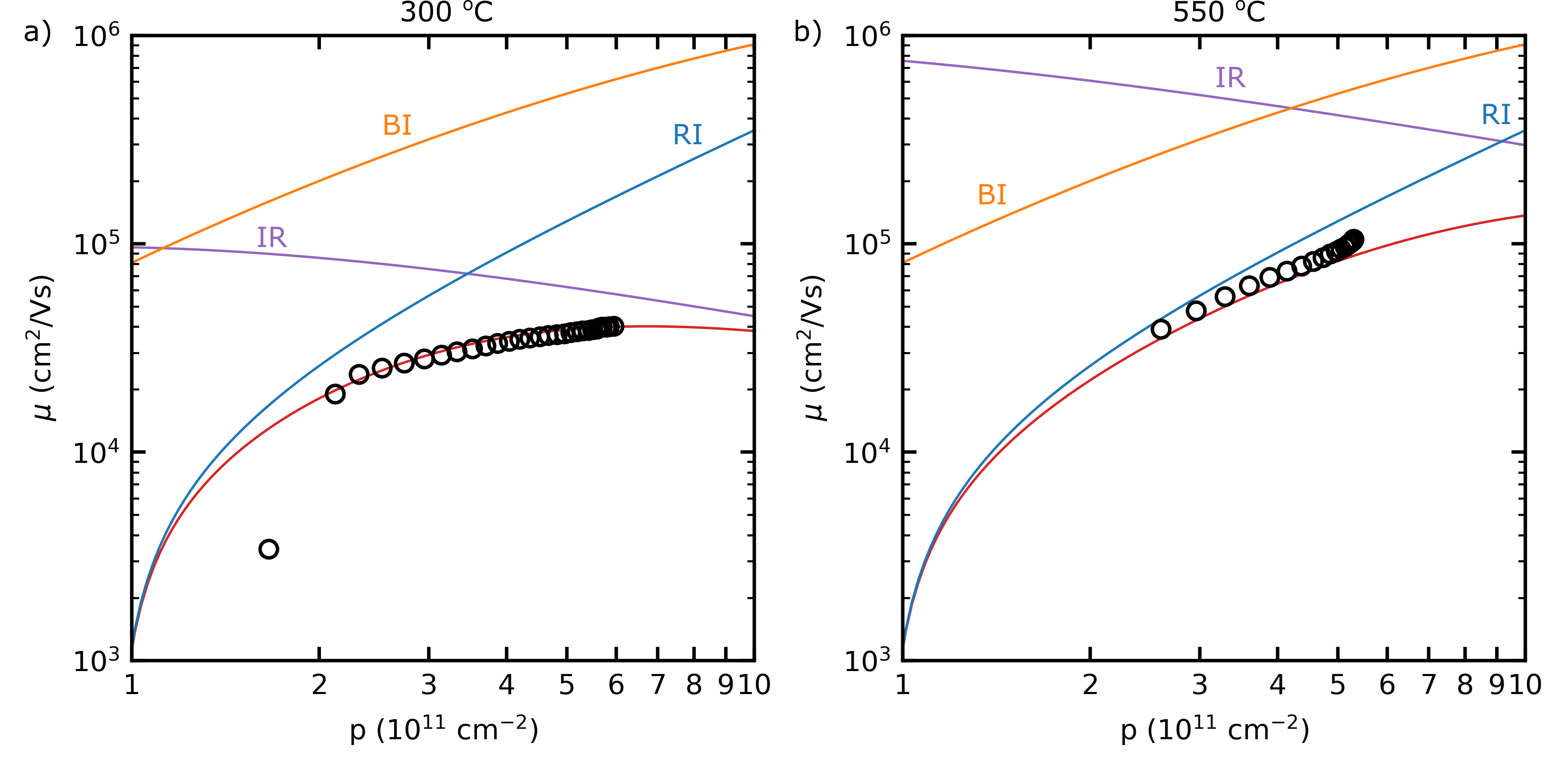}
\caption{\label{fig:6} Experimentally measured mobility and calculated mobility at varying carrier densities for the Ge QWs grown with a) T\textsubscript{QW} = 300 \unit{\degreeCelsius} and b) T\textsubscript{QW} = 550 \unit{\degreeCelsius}. The scattering from remote ionized impurities (RI), interface roughness (IR), and background impurities (BI) was considered in the mobility model. }
\end{figure*}

To understand the differences in mobility for the low temperature and high temperature growth and how to further improve the mobility of the samples, the mobility of the Ge QWs was modeled using the formalism introduced by Stern and Howard \cite{Stern1967}. The scattering times of the holes in the Germanium quantum well were numerically calculated. Scattering from remote impurities at the semiconductor top surface, background impurities distributed throughout the heterostructures, and interface roughness was considered for modeling the mobility of the Ge QWs. Negligible scattering from threading dislocation is expected at the observed threading dislocation densities \cite{Debdeep2000}, and the scattering from threading dislocations was neglected. The scattering times were computed independently then the total scattering time was calculated using Matthiessen’s rule. A non-parabolic effective mass of the Ge QW was used for the calculation, which was solved for using the $k \cdot p$ method \cite{Vurgaftman2021}. The exact details of the mobility modeling are given in \cite{Dong2024prm}, with modifications to the scattering potentials to account for the undoped heterostructures and multiple scattering events \cite{fang_negative_1966,gold_mobility_1989,Gold1991,gold_metalinsulator_2011,Gold2013}.

The measured mobility and the modeled mobility for the Ge QWs grown at T\textsubscript{QW} = 300 \unit{\degreeCelsius} and T\textsubscript{QW} = 550 \unit{\degreeCelsius} are shown in Fig. \ref{fig:6}. The mobility limits due to the different scattering mechanisms are also plotted. A background impurity density of $5\times10^{15}$ cm\textsuperscript{-3} and a remote impurity density of $2.3 \times 10^{12}$ cm\textsuperscript{-2} were used for both samples.  Interface roughness scattering of charge carriers is sensitive to both the interface roughness as well as the the roughness correlation length. Roughness correlation lengths comparable to the Fermi wavevector are known to enhance the effects of interface roughness and reduce the mobility \cite{Sakaki1987}. The measured AFM roughness values were used as the interface roughness, while the roughness correlation lengths were subsequently varied to obtain good agreement between the experimental results and the measured results. The mobility of the Ge QW grown at 300 \unit{\degreeCelsius} was modeled with an interface roughness of 1.3 nm with a roughness correlation length of 85 nm, while the Ge QW grown at 550 \unit{\degreeCelsius} was modeled with an interface roughness of 1.3 nm with a roughness correlation length of 155 nm. 

In both samples, good agreement between the measured mobility and calculated mobility is obtained. Remote impurity scattering from impurities at the semiconductor top surface limit the mobility at low carrier densities, while the scattering from background impurities does not limit the mobility in these samples. For the sample grown at T\textsubscript{QW} = 300 \unit{\degreeCelsius}, interface roughness scattering appears to limits the mobility at higher carrier densities and results in the weak dependence of mobility on carrier density. For the Ge QW grown at T\textsubscript{QW} = 550 \unit{\degreeCelsius}, the improved mobility and increased slope of the dependence of mobility with carrier density is likely due to a reduction in the interface roughness scattering, which is in turn due to an increase in the roughness correlation length. The perceived increase in roughness correlation length is attributed to the higher temperature growth promoting adatom diffusion and longer roughness lengths scales, as observed in the AFM micrographs. Further transmission electron microscopy experiments could quantify the changes in roughness correlation length in the various samples, but this experiment is beyond the scope of this study. In the Ge QW grown at T\textsubscript{QW} = 550 \unit{\degreeCelsius}, interface roughness scattering is no longer the apparent limiting scattering mechanism at higher carrier densities, and remote impurity scattering instead remains the limiting scattering mechanism. To further improve the mobility of the highest quality samples to values equivalent to the highest reported mobilities in Ge QWs of similar depth from the surface, these results suggest that improvements in the device processing to reduce the density of remote impurities are necessary.

\section*{Conclusions}

In conclusion, the growth of Ge QWs grown by MBE was investigated. Thick RGBL are required in order to reduce the defect density and eliminate line-shaped defects. The QW growth temperature was optimized, with the optimal growth temperature for maximizing mobility being the highest growth temperature before roughening of the film occurs. The mobility modeling of the Ge QWs suggests that higher growth temperatures promote longer interface roughness correlation lengths, which in turn correlates with reduced interface roughness scattering in the Ge QWs. In the Ge QWs with 22-nm-thick top spacer grown in this study, a maximum 2 K mobility of 105,000 cm\textsuperscript{2}/Vs is obtained. A narrow growth temperature window for maximizing mobility is observed. The highest mobility samples are limited by scattering from impurities at the top surface. To further improve the mobility of the MBE-grown samples, improved device processing is required to reduce the surface state density at the top surface.

%\section*{Acknowledgments}

%TBD

\section*{Data Availability}
The data that support the findings of this study are available from the corresponding author upon reasonable request.

\section*{References}

\bibliography{References}

\end{document}